\documentclass[aps,prl,reprint,superscriptaddress]{revtex4-2}
\usepackage[colorlinks,linkcolor=blue,hyperindex,CJKbookmarks]{hyperref}
\usepackage{physics}
\usepackage{graphicx}

\begin{document}

\title{Determinantal Quantum Monte Carlo solver for Cluster Perturbation Theory}
\author{Edwin W. Huang}
\email{edwinwhuang@gmail.com}
\affiliation{Department of Physics and Institute for Condensed Matter Theory, University of Illinois at Urbana-Champaign, Urbana, IL 61801, USA}
\author{Yao Wang}
\email{yaowang@g.clemson.edu}
\affiliation{Department of Physics and Astronomy, Clemson University, Clemson, SC 29631, USA}
\begin{abstract}
Cluster Perturbation Theory (CPT) is a technique for computing the spectral function of fermionic models with local interactions. By combining the solution of the model on a finite cluster with perturbation theory on intra-cluster hoppings, CPT provides access to single-particle properties with arbitrary momentum resolution while incurring low computational cost. Here, we introduce Determinantal Quantum Monte Carlo (DQMC) as a solver for CPT. Compared to the standard solver, exact diagonalization (ED), the DQMC solver reduces finite size effects through utilizing larger clusters, allows study of temperature dependence, and enables large-scale simulations of a greater set of models. We discuss the implementation of the DQMC solver for CPT and benchmark the CPT+DQMC method for the attractive and repulsive Hubbard models, showcasing its advantages over standard DQMC and CPT+ED simulations.
\end{abstract}

\maketitle

\textbf{Introduction:}
The spectral function $A(k,\omega)$ is a fundamental quantity in many-body physics. A model's spectral function directly reflect the properties of its elementary charged excitations and is useful for characterizing both ordered and unordered phases. Experimentally, the spectral function can by measured by angle-resolved photoemission and is directly related to the density of states measured in tunneling spectroscopy\,\cite{damascelli2003angle}. These experimental probes have been instrumental in characterizing electronic structure and phase diagrams. Continual improvements in energy and momentum resolution have led to significant advances in our understanding of the properties of quantum materials\,\cite{Sobota2021}.

As a dynamical quantity, the selection of techniques for calculating the spectral function in models of interacting electrons is limited. The most common methods can be classified roughly into three categories: perturbative methods, finite-cluster methods, and embedding methods. Perturbative methods (e.g.~\cite{Bickers1989,tremblay2006}) have the advantage of being computationally inexpensive and therefore can be applied to large clusters, leading to fine momentum resolution and minimal finite-size effects. However, their validity in intermediate and strongly interacting models is questionable at best. Finite cluster methods, such as exact diagonalization (ED)\,\cite{li1991spectral,dagotto1992single}, determinantal Quantum Monte Carlo (DQMC)\,\cite{bss1981,white1989}, and density matrix renormalization group (DMRG)\,\cite{white1992,White2004}, treat interactions exactly, but ultimately have limitations that manifest as a restriction in the accessible system size. Embedding methods seek to treat the model on an infinite lattice, by solving an `embedded' finite cluster with an exact method and treating longer-range correlations approximately. This allows for continuous momentum resolution, with finite-size effects that can be controlled by increasing the size of the embedded cluster\,\cite{maier2005quantum}. In this Letter we focus on one such embedding method, cluster perturbation theory (CPT)\,\cite{pairault1998strong,Senechal2000,Senechal2002,senechal2012book}, and introduce DQMC as a novel solver for CPT. After introducing the formalism and implementation of the CPT+DQMC method, we will explore three examples: the attractive Hubbard model in a superconducting state, the half-filled repulsive Hubbard model, and the doped repulsive Hubbard model. In these examples, we will illustrate the advantages of the DQMC solver over the standard ED solver for CPT\,\cite{gros1993cluster,Senechal2000, Senechal2002, senechal2004hot, ning2006phonon, Kohno2012, Kuzmin2014,kohno2014spectral,wang2015origin,kohno2015spectral,wang2018influence, kuz2019effect,kuz2020doping,wang2020emergence}, and also demonstrate the advantages of the CPT+DQMC method over finite size DQMC simulations. We conclude with a brief discussion of interesting open problems that are particularly suitable for study by CPT+DQMC.

\begin{figure}
\includegraphics[width=\columnwidth]{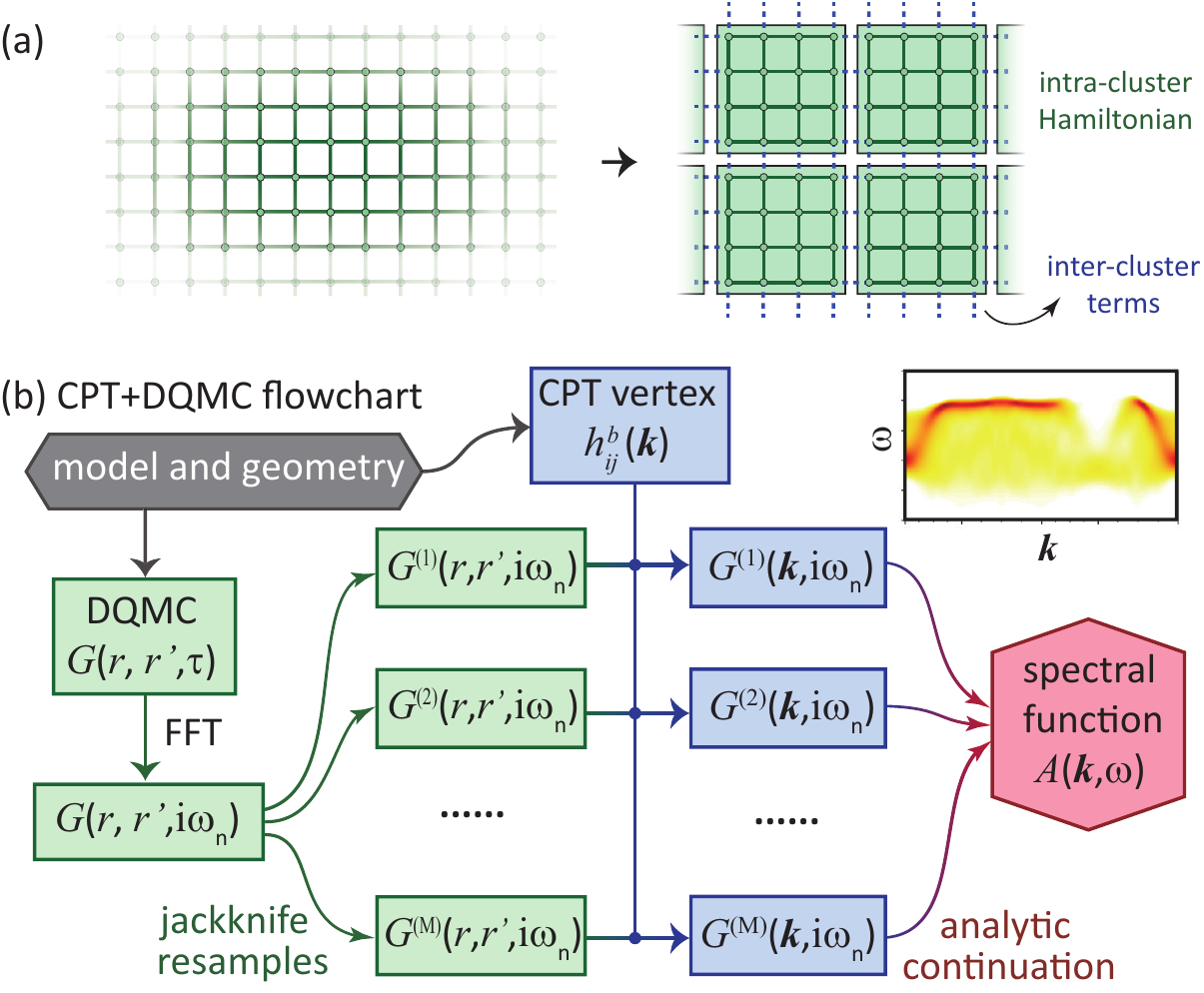}
\caption{(a) Schematic of CPT method: the infinite plane is divided into clusters. The Green's function associated with each cluster Hamiltonian is solved exactly, while the inter-cluster terms are treated as perturbation. (b) Flowchart of the DQMC+CPT method.}
\label{fig:flowchart}
\end{figure}

\textbf{CPT formalism:}
As the CPT formalism has been derived and discussed in detail in Refs.~\cite{Senechal2000,Senechal2002,maier2005quantum}, we will only summarize the most important results below. Consider an infinite lattice that can be separated into clusters of $N^c$ sites. CPT applies to Hamiltonians involving local interactions, such that only the hopping terms connect different clusters [see Fig.~\ref{fig:flowchart}(a)]. Such a Hamiltonian can be decomposed as
\begin{equation}
\mathcal{H} = \sum_{\mathcal{C}} \mathcal{H}^\mathcal{C} + \sum_{i j \sigma}  h^{b}_{i j} c_{i \sigma}^\dagger c_{j \sigma}.
\end{equation}
Here, $\mathcal{H}^\mathcal{C}$ contains all the hopping and interaction terms involving only a single cluster $\mathcal{C}$. The inter-cluster hoppings are contained in the matrix $\mathbf{h}^{b}$. For simplicity, we consider a single-orbital problem, although the generalization of the formalism to multi-orbital models is straightforward. With the intra- and inter-cluster terms of the Hamiltonian separated, the Green's function in CPT is given by
\begin{equation}
\mathbf{G}(z) = \mathbf{G}^c(z) \left[\mathbf{I} - \mathbf{h}^b \mathbf{G}^c(z)\right]^{-1}, \label{eq:cpt_real}
\end{equation}
where $\mathbf{G}^c(z)$ is the Green's function calculated at complex frequency $z$ using the cluster Hamiltonian $\mathcal{H}^\mathcal{C}$. The bold symbols denote matrices in the real-space basis. Clearly, $\mathbf{G}^c(z)$ is block-diagonal, with identical blocks of size $N^c \times N^c$ (these blocks can be different when a supercluster is employed). To calculate $\mathbf{G}^c(z)$ requires solving $\mathcal{H}^\mathcal{C}$ i.e.~the model on a cluster of $N^c$ sites with open boundary conditions. The original and most commonly used CPT algorithms employ ED as the cluster solver, which was extended to tDMRG recently\,\cite{Yang2016}. As we will show, DQMC is also suitable as a solver and exhibits its advantage in particular problems.

Strictly speaking, we seek to evaluate
\begin{equation}
G(k,z) = \frac{1}{N^c} \sum_{r, r'} \sum_{R'} e^{-i k r} e^{i k (r' + R')} G_{r, r' + R'}(z),
\end{equation}
where $r$ and $r'$ are in one cluster, and $R'$ connects equivalent sites in different clusters. For a given momentum $k$, the matrix algebra of Eq.~\eqref{eq:cpt_real} can be simplified in order to involve only $N^c \times N^c$ matrices. The final result is
\begin{equation}
G(k,z) = \frac{1}{N^c} \sum_{r,r'}e^{-i k (r - r')} \qty[\frac{\mathbf{G}^c(z)}{\mathbf{I} - \mathbf{\Tilde{h}}^b \mathbf{G}^c(z)}]_{r,r'}. \label{eq:cpt_k}
\end{equation}
Here, all matrices are $N^c \times N^c$ and $\mathbf{\Tilde{h}}^b$ is defined as
\begin{equation}
\Tilde{h}_{r,r'}^b = \sum_{R'} e^{i k R'} h_{r,r' + R'}^b.
\end{equation}
Finally, the spectral function is defined as $A(k, \omega) = \frac{-1}{\pi} \Im G(k, \omega + i 0^+)$.

\textbf{CPT+DQMC implementation:}
DQMC is a numerically exact algorithm for simulating interacting systems on finite clusters at finite temperature\,\cite{bss1981,white1989}. Utilizing DQMC as the cluster solver in CPT and obtaining the spectral function involves the sequence of operations sketched in Fig.~\ref{fig:flowchart}(b). The details of the operations are as follows:

\begin{enumerate}
\item DQMC is used to simulate a cluster with $N^c$ sites with open boundary conditions. Unlike in simulations with periodic boundaries, translation symmetry must not be applied to the measurements. The output of the DQMC calculation is the imaginary time Green's function
\begin{equation}
G_{r r'}^c(\tau) = -\ev{c_{r \sigma}(\tau) c_{r' \sigma}^\dagger},
\end{equation}
where $i$ and $j$ are site indices. We assume spin rotation symmetry and hence omit the spin index $\sigma$.
\item The data are Fourier transformed to Matsubara frequencies
\begin{equation}
G_{r r'}^c(i \omega_n) = \int_0^\beta \dd{\tau} e^{i \omega_n \tau} G_{r r'}^c(\tau).    
\end{equation}
Because imaginary time is discretized in DQMC, the integral over $\dd{\tau}$ must be evaluated carefully to avoid inaccuracy at large Matsubara frequency. Our approach is to interpolate $\tau$ with cubic splines onto a very fine grid before integrating numerically. Since both the interpolation and the Fourier transform are linear operations, they may be combined into a single matrix. A single matrix multiplication is used to perform these operations on all $r, r'$ and all bins. Since $G_{r r'}^c(i \omega_n)$ is complex, the number of Matsubara frequencies kept needs to be only half the number of imaginary time points for a 1-to-1 transformation.
\item The bins of $\mathbf{G}^c(i \omega_n)$ data are resampled by either jackknife or bootstrap resampling. If the model has a fermion sign problem, the average sign is divided in this step.
\item The resampled data is combined with the inter-cluster hopping through Eq.~\eqref{eq:cpt_k} to calculate the CPT Matsubara Green's function $G(k, i\omega_n)$. As this is done for every resample and may be slightly time consuming, it is advantageous to focus on $k$ along high-symmetry cuts.
\item The spectral function is extracted from the relation
\begin{equation}
G(k, i\omega_n) = \int \dd{\omega} \frac{A(k, \omega)}{i\omega_n - \omega}
\end{equation}
by numerical analytic continuation, for every $k$ point. In the examples shown later, we use the Maximum Entropy Method (MaxEnt)\,\cite{Jarrell1996} with a flat model function and the prescription of Ref.~\cite{Bergeron2016} for choosing the entropy weight. Note that since the resampled data are not independent, the estimated covariance matrix must be multiplied by a correction factor. For jackknife resampling, this factor is $(M-1)^2$ where $M$ is the number of bins.
\end{enumerate}

The specific order of these steps is due to the following constraints. First, most methods of analytic continuation, including MaxEnt, assume a non-negative spectral function. Because the sign of $A_{i j}(\omega)$ is frequency dependent for $i \neq j$, CPT and the transformation to momentum space should be applied before analytic continuation. Second, CPT is specified in frequency space. Together with the first constraint, this necessitates the use of Matsubara frequencies. Finally, the Green's function in CPT is a non-linear function of $\mathbf{G}^c$, so the sampling noise should be minimized by jackknife or bootstrap resampling before applying Eq.~\eqref{eq:cpt_real}. Steps 2 and 3 may be reordered because both resampling and the Matsubara transformation are linear.

\textbf{Application to the Hubbard model:}
The Hubbard model Hamiltonian is
\begin{equation}
\mathcal{H} = -t \sum_{\ev{i j}} \left(c_{i \sigma}^\dagger c_{j \sigma} + h.c.\right) + U \sum_i n_{i \uparrow} n_{i \downarrow},
\end{equation}
where $t$ is the hopping between nearest neighbors $i$ and $j$, and $U$ is the interaction strength. We will consider both the attractive Hubbard model with $U/t = -4$ and the repulsive model with $U/t = 8$. In all cases, we consider the 2D square lattice and use square cluster geometries. Typically, for each simulation we run $\sim 1000$ Markov chains with $\sim 10^6$ sweeps each, and measure $G(\tau)$ every other sweep. More measurements are collected for simulations with a severe sign problem. The standard error in $G(\tau)$ is $\sim 10^{-5}$, which is low enough that the MaxEnt analytic continuation is highly repeatable.

We will first consider two simple cases of the Hubbard model that are well understood and free of the sign problem: the doped attractive Hubbard model and the half-filled repulsive Hubbard model. The absence of a sign problem allows for DQMC simulations on large clusters and low temperatures. Therefore, we will use large-cluster DQMC simulations as a reference to evaluate the performance of CPT+DQMC simulations, which involve much smaller clusters.

\begin{figure}
\includegraphics{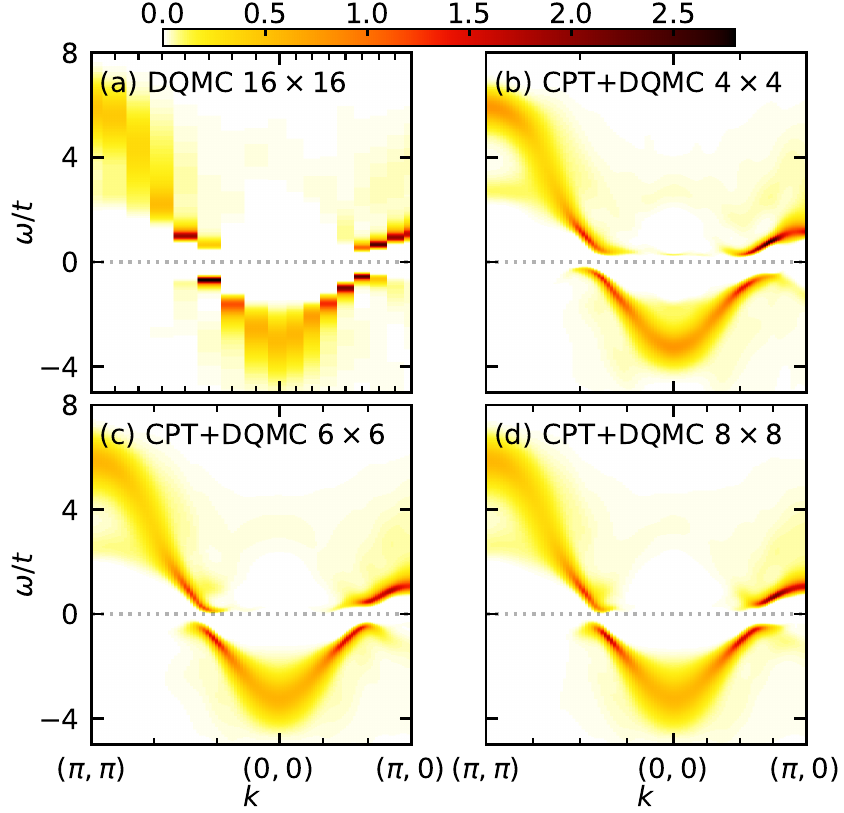}
\caption{Spectral function $A(k,\omega)$ of of the attractive Hubbard model calculated by DQMC on a $16 \times 16$ cluster (a) and CPT+DQMC on clusters of sizes as labelled (b-d). The interaction strength is $U/t = -4$, average filling is $\ev{n} \approx 0.63$, and the temperature is $T/t = 1/12$.}
\label{fig:negU4}
\end{figure}

The attractive Hubbard model is an $s$-wave superconductor upon doping. At $U/t = -4$ and an average filling $\ev{n} \approx 0.6$, the critical temperature has been determined to be $T_c/t \approx 0.15$ from finite-size scaling of DQMC simulations\,\cite{Paiva2004}. These parameters are close to optimal, in the sense of maximizing $T_c$. In Fig.~\ref{fig:negU4}, we plot the spectral function of the attractive Hubbard model with these parameters at a temperature $T/t = 1/12$ well under the nominal superconducting transition. In the $16 \times 16$ DQMC simulation of Fig.~\ref{fig:negU4}(a), we see a particle-hole symmetric superconducting gap separating sharp Bogoliubov quasiparticle peaks. Back-bending of the dispersion is visible but the linewidth increases rapidly when moving away from the Fermi momenta, since the solution involves correlations beyond the mean-field.

In the spectra computed by CPT+DQMC [see Fig.~\ref{fig:negU4}(b-d)], the broad high-energy features are essentially identical to those from DQMC in Fig.~\ref{fig:negU4}(a). The low-energy features are more revealing, and demonstrate both the advantages and limitations of CPT. In the $4 \times 4$ CPT+DQMC spectra of Fig.~\ref{fig:negU4}(b), although the back-bending dispersions are very clear, the superconducting gap appears to be indirect and particle-hole asymmetric. These anomalies are undoubtedly finite-size artifacts, which are not completely corrected by CPT, and we find that they diminish when increasing the cluster size to $6 \times 6$ and $8 \times 8$ (Fig.~\ref{fig:negU4}(c-d)). This highlights the importance of systematically checking cluster size dependence and illustrates the challenges of studying spectral features associated with superconductivity by CPT+ED simulations, which are limited to $\sim 20$ sites.

\begin{figure}
\includegraphics{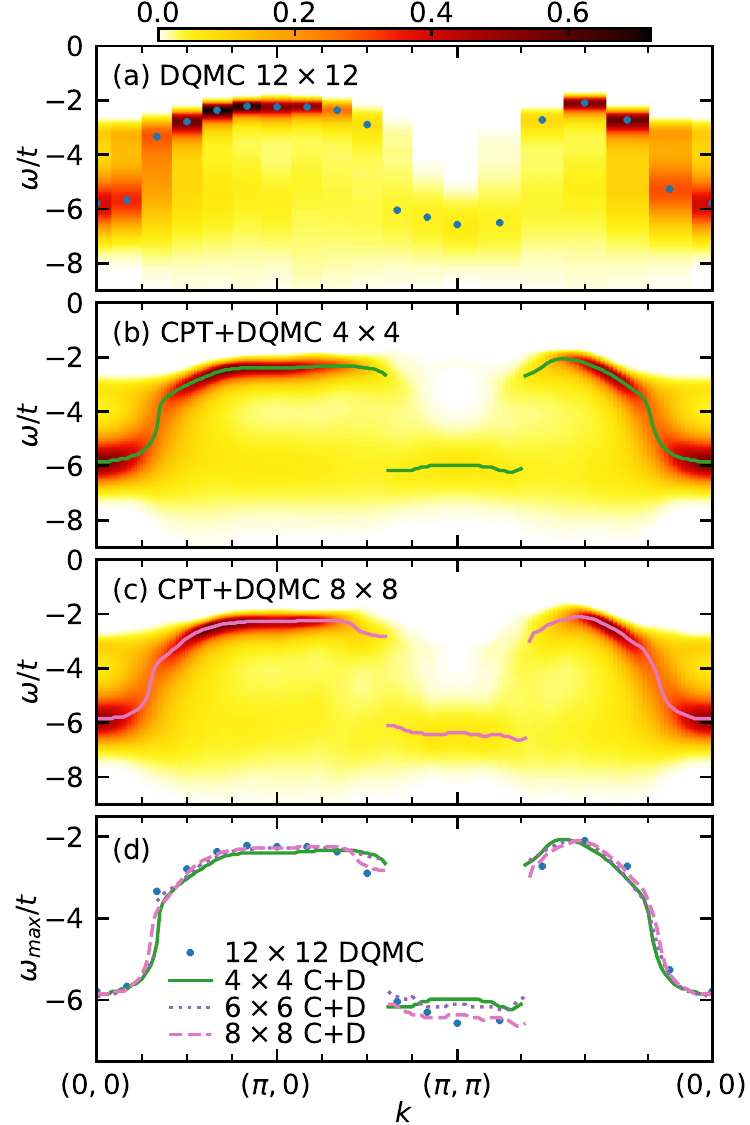}
\caption{Spectral function $A(k,\omega)$ of the half-filled Hubbard model calculated by DQMC on a $12 \times 12$ cluster (a), DQMC+CPT on a $4 \times 4$ cluster (b), and DQMC+CPT on a $8 \times 8$ cluster (c). Energy distribution curve (EDC) peaks are indicated by dots in (a), indicated by solid lines in (b) and (c), and compared in (d). The interaction strength is $U/t = 8$ and the temperature is $T/t = 1/16$.}
\label{fig:U8mu0}
\end{figure}

The repulsive Hubbard model at $U/t=8$ and half-filling is an antiferromagnetic (AFM) Mott insulator. We plot its spectral function in Fig.~\ref{fig:U8mu0} for a temperature $T/t = 1/16$\,\cite{removalFootnote}. Comparing Fig.~\ref{fig:U8mu0}(a) and (b) shows that CPT+DQMC on a $4 \times 4$ cluster produces highly accurate spectra qualitatively identical to that from the larger $12 \times 12$ DQMC calculation. Interestingly, the $4 \times 4$ CPT+DQMC demonstrates a slight momentum asymmetry in the removal spectrum's dispersion around $(\frac{\pi}{2}, \frac{\pi}{2})$. Due to the strong AFM of the system, dispersions should be symmetric about the AFM zone boundary, and the slight asymmetry in the $4 \times 4$ CPT+DQMC is likely a consequence of the limited range of AFM correlations in a $4 \times 4$ cluster. Indeed, we find that in a larger $8 \times 8$ CPT+DQMC calculation as shown in Fig.~\ref{fig:U8mu0}, the EDC peak dispersion near $(\frac{\pi}{2}, \frac{\pi}{2})$ is almost symmetric.


\begin{figure}
\includegraphics{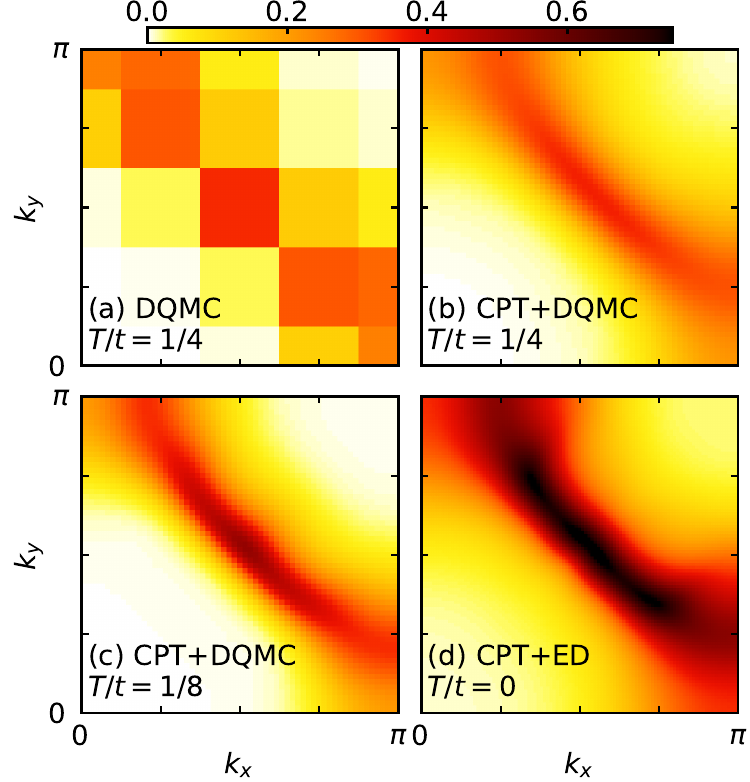}
\caption{Low energy spectral weight $A(k,\omega=0)$ of the hole-doped Hubbard model calculated by (a) DQMC on a $8 \times 8$ cluster at $T/t = 1/4$, (b) CPT+DQMC at $T/t = 1/4$, (c) CPT+DQMC at $T/t = 1/8$, and (d) CPT+ED at $T=0$ with a broadening of $0.15t$. All CPT calculations used $4 \times 4$ clusters. The interaction strength is $U/t = 8$ and the average filling is $\ev{n} \approx 0.94$.}
\label{fig:U8doped_fs}
\end{figure}

Having tested CPT+DQMC in two models that are well understood, we now explore the suitability of CPT+DQMC for the less understood case of the doped repulsive Hubbard model. Most progress in understanding the model has come from ground state calculations of static observables, and it has been fruitful to compare calculations across many methods involving different approximations\,\cite{Leblanc2015}. However, far fewer methods are capable of calculating the spectral function, a momentum-resolved dynamical quantity. In the doped model, DQMC is limited by the sign problem to temperatures $T/t \gtrsim 0.2$ \cite{Loh1990,Iglovikov2015}, and most calculations of $A(k,\omega)$ at lower temperatures are based on cluster extensions or variants of dynamical mean field theory\,\cite{maier2005quantum,tremblay2006}.

In Fig.~\ref{fig:U8doped_fs}, we show the spectral function at zero frequency of the Hubbard model at $6\%$ hole doping, calculated by DQMC, CPT+DQMC, and CPT+ED. In the $8\times 8$, $T/t=1/4$ simulation of Fig.~\ref{fig:U8doped_fs}(a), the average fermion sign is $0.049$.
The fact that CPT+DQMC is capable of continuous momentum resolution by simulating small, open-boundary clusters is an enormous advantage: the average sign of the $4\times4$ simulations in Fig.~\ref{fig:U8doped_fs}(b) is $0.70$, and the continuous momentum dependence reveals clearly that the underlying Fermi surface is hole-like. In fact, $4\times4$ CPT+DQMC simulations can be pushed to $T/t=1/8$ where the average sign is $0.075$. As evident in Fig.~\ref{fig:U8doped_fs}(c), and consistent with the zero temperature CPT+ED calculation in Fig.~\ref{fig:U8doped_fs}(d), the difference in intensity between the nodal and anti-nodal directions (nodal-antinodal dichotomy) is very strong, and is likely related to the pseudogap.

\textbf{Discussion:}
We have implemented and demonstrated CPT+DQMC as a powerful method for calculating the spectral function in correlated electron systems. By using DQMC as a solver, the accessible cluster sizes are significantly larger than possible by CPT+ED. We have shown in Figs.~\ref{fig:negU4} and \ref{fig:U8mu0} some of the benefits of these larger cluster sizes. In general, we find that $4 \times 4$ CPT simulations are highly accurate apart from low-energy details related to ordering or long-range correlations. One unexplored possibility is to develop novel CPT schemes that account for the presence of ordering.

Compared to standard DQMC simulations, CPT+DQMC has the advantage of achieving continuous momentum resolution and relatively high accuracy with a small cluster and lower computational cost. This advantage has been demonstrated and employed in ED+CPT simulations, as an efficient correction for the finite-size effect in pure ED. In models with a sign problem, such as the doped Hubbard model studied in Fig.~\ref{fig:U8doped_fs}, the advantage of DQMC+CPT potentially allows access to parameter regimes inaccessible by moderate-size DQMC simulations. Even for sign-free models, the size of DQMC simulations required to resolve features such as dispersion back-bending (Fig.~\ref{fig:negU4}) is considerable, and hence the continuous momentum resolution of CPT+DQMC is highly beneficial.

Perhaps the most important opportunity enabled by CPT+DQMC is the possibility to study models that are not well suited to ED. For instance, strongly electron-phonon coupled system involves huge Hilbert space and cannot be solved by pure ED\,\cite{nath2016phonon,nath2015interplay,payeur2011variational, wang2016using, wang2018light}. However, both the Hilbert-space issue and the fermion-sign issue are absent for DQMC in the electron-phonon systems. An exciting prospect would be to use CPT+DQMC to investigate the spectral signatures of the breakdown of Eliashberg theory in the Holstein model\,\cite{Esterlis2018}, which has only local interactions and local phonon degrees of freedom. Similarly, multi-orbital models are less problematic for DQMC than for ED, and there is a considerable phase space of sign-free multi-orbital models with rich phase diagrams\,\cite{Wu2005,Huang2020}. We look forward the the application of CPT+DQMC to these systems.

\begin{acknowledgements}
EWH was supported by the Gordon and Betty Moore Foundation EPiQS Initiative through the grants GBMF 4305 and GBMF 8691. Y.W. acknowledges support from the National Science Foundation (NSF) award DMR-2038011. The calculations were performed on the Frontera computing system at the Texas Advanced Computing Center.
\end{acknowledgements}

\bibliographystyle{apsrev4-2}
\bibliography{main}

\end{document}